# A Review-based Taxonomy for Secure Health Care Monitoring: Wireless Smart Cameras

Ravi Teja Batchu[1], Abeer Alsadoon*[1,2,3,4], P.W.C. Prasad[1,5], Rasha S. Ali[6], Tarik A. Rashid[7], Ghossoon Alsadoon[8], Oday D. Jerew[4]

[1]School of Computing and Mathematics, Charles Sturt University (CSU), Wagga Wagga, Australia
[2]School of Computer Data and Mathematical Sciences, University of Western Sydney (UWS), Sydney, Australia
[3]Kent Institute Australia, Information Technology Department, Sydney, Australia
[4]Asia Pacific International College (APIC), Information Technology Department, Sydney, Australia
[5]Australian Institute of Higher Education, Sydney, Australia
[6]Department of Computer Techniques Engineering, AL Nisour University College, Baghdad, Iraq
[7]Computer Science and Engineering, University of Kurdistan Hewler, Erbil, KRG, Iraq
[8]Business Informatics Department, AMA International University Bahrain (AMAIUB)

**Abeer Alsadoon[1*]**
* Corresponding author. Dr. Abeer Alsadoon, School of Computing and Mathematics, Charles Sturt University (CSU), Wagga Wagga, Australia, Email: alsadoon.abeer@gmail.com , Phone +61413971627

## Abstract

Health records data security is one of the main challenges in e-health systems. Authentication is one of the essential security services to support the stored data confidentiality, integrity, and availability. This research focuses on the secure storage of patient and medical records in the healthcare sector where data security and unauthorized access is an ongoing issue. A potential solution comes from biometrics, although their use may be time-consuming and can slow down data retrieval. This research aims to overcome these challenges and enhance data access control in the healthcare sector through the addition of biometrics in the form of fingerprints. The proposed model for application in the healthcare sector consists of Collection, Network communication, and Authentication (CNA) using biometrics, which replaces an existing password-based access control method. A sensor then collects data and by using a network (wireless or Zig-bee), a connection is established, after connectivity analytics and data management work which processes and aggregate the data. Subsequently, access is granted to authenticated users of the application. This IoT-based biometric authentication system facilitates effective recognition and ensures confidentiality, integrity, and reliability of patients' records and other sensitive data. The proposed solution provides reliable access to healthcare data and enables secure access through the process of user and device authentication. The proposed model has been developed for access control to data through the authentication of users in healthcare to reduce data manipulation or theft.

**Keywords:** Healthcare sector, biometric technique, access control

# Introduction

Data management in the medical domain is designed to secure and monitor healthcare data in the form of patient records as well as disease and treatment information. Presently, access to health data is protected through biometrics at different access levels. While this system provides some protection, frequently reported instances of data manipulation and theft indicate that there are still challenges, one of which is also an inability to authenticate large volumes of data [1].

The system proposed in this research adds Raspberry Pi with SlugCam and machine learning biometric authentication based on a wireless camera or a sensor network that will store large amounts of multidimensional data [2]. The current literature on biometrics was reviewed to identify authentication techniques to provide access control to health data with a focus on all-inclusive systems that can predict vulnerabilities and unauthorized access [3]. Component classification, therefore, was a significant aim of this review [4]. It was found that current systems cannot provide secure access to multi-dimensional data or detect unauthorized access to the system based on data and samples [5]. A model is, therefore, proposed to overcome these issues using a system built with a raspberry pi-based slug camera with machine learning [6]. Only current research that offers biometric access control techniques based on wireless networks is presented by Alsmirat et. [1]. However, it does provide high-level access control and authentication [7]. Component verification is based on accuracy, fit [8] and completeness [9] [10].

The work is divided as follows: Section 2 contains the literature review. In section 3, the state-of-the-art model is discussed along with components. In thefourth section, the proposed model is presented together with a detailed discussion of the solutions it offers to the limitations of the state-of-the-art model. In section 5, validation and evaluation are performed, and section 6 verifies the proposed technique, followed by sections 7 and 8 with discussion and conclusion.

*:* Raspberry Pi with SlugCam and machine learningallows multi-dimensional records to be stored at high definition. It is effective in evaluating security and provides access control. The basis of the work is the inclusion of healthcare data, classifications, access control policies, feature extraction, and raspberry pi. The primary components of the system are collection, network communication, and authentication.

## Biometric-based security techniques

Several papers were reviewed, each providing models or systems representing authentication techniques focused on at least one of the issues of importance in health care, namely security, access control and authenticity [11] [12]. A gap was identified in existing research in terms of biometric authentication techniques suitable for the healthcare sector, where the need for security and authentication is high [13]. None considered the aspect of healthcare monitoring and authorization prediction of users. *Biometric healthcare system:* Hamidi [14] designed an IoT-based biometric technique for the healthcare system that preserved data integrity and assured data availability, accessibility and security. The biometric technique identified physical characteristics embedded in a Smart system. The limitations of this system lie in its ambitious combination of biometrics. From a range of possible biometric feature choices, they identified fingerprinting and keystrokes as two that would need to be combined to provide the right level of security. However, the author acknowledged that such a system would incur cost probably beyond what even healthcare providers are willing to invest. The paper was, nevertheless, of interest due to the inclusion of a machine learning algorithm. *Priority-based parallel algorithm:* Shakil et al.introduced a biometric authentication technique in which the access hierarchy is defined. At the enrolment phase, a priority value is assigned [15]. If this is 4 on a scale of 1-4, then the security level assured by the system is the maximum. Biometric signatures ensure the authenticated access to e-healthcare data to manage secure access. This system is of interest as it provides an example of efficient biometric authentication although it does not address health care monitoring. *Lightweight key agreement protocol with rekeying:* Meena et al. , used an algorithm for providing energy-efficient secure transmission of data used for security monitoring, basing their security regime on a DTSL handshake protocol. The limitations imposed by medical sensors were overcome by interposing an intermediary system for authentication of the sensor input[16]. This research is valuable because it offers an energy-efficient secure transmission that overcomes the sensor resource constraint.

Pirbhulal et al. (2018) offered an energy-efficient fuzzy vault based biometric security approach that increases security and authenticity[17]. A biometrics-based security mechanism is said to lead to enhanced security and EFVSM reduces extensive energy usage through time-domain analysis using EI Fuzzy vault. This achieves data consistency and enhanced security. The limitation is that the research has not focused on securing communication which is also necessary for data consistency and secure transmission. *Elliptic curve cryptography (ECC) based secure three-factor user authentication :* Challa et al.have reduced communication and computation cost through performing a password and biometric update externally without the involvement of third parties. This enhances security aspects[18]. The limitation of this work is that security was not analyzed through an informal security analysis for verification [19]Nevertheless, this research is valuable as security is enhanced through an ECC based three-factor authentication approach, a smart card, user biometrics and user passwords. An expansion of the model is planned for future work. *Formal authentication techniques:* Amin et al. contribute by preserving the anonymity of users and enhance the performance of the patient monitoring system by managing its authentication and security aspects[20]. The effectiveness of the patient monitoring system secures the authentication of users. However, energy optimization was not addressed, limiting this work, although the robustness of the patient monitoring system [is of interest for the current work [14].

Furthermore, Qi et al. provided information about the existence of a crucial agreement protocol to secure communication between wearable sensors and medical professionals[21]. As per the study, the end-to-end authentication protocol prevents unauthorized access and never refuses access for any authenticated participant[22]. The pitfall is that the research focuses on authenticity and security, but the effectiveness of monitoring health was not addressed. Still, the work is valuable for protecting healthcare information and grants secure access to healthcare applications. *Compressed sensing (CS)-based security mechanism:* Nandhini et al. have focused on security aspects and also enabled effective resource utilization [23]. They have further reduced storage complexity by applying security keys to only the significant video frames without compromising the desired security level. However, this technology has an issue related to energy consumption and storage requirements. A solution has come from Hussein et al., who addressed the problem associated with the use of different measurement matrices for security and compression [24]. Therefore, the advancement made was to reduce storage needs by 92% and energy consumption by 53%. The given solution will use a single measurement matrix for the compression as well as security. The limitation is that when the distance between the sensor node and destination node increases, recovering the lost packet is reduced. This work is, nevertheless, valuable for increasing security by preventing attackers from reconstructing the video[25].

Furthermore, Wang et al. enhanced the tracking quality and contributed to the reduction of the moving distance using wireless camera sensor networks [26]. There are issues with this solution in terms of sensing range and distance for tracking targets, which affect the security aspects. This negatively affects the consumption of energy of mobile devices used for the purpose of tracking. Quality tracking was also developed by Jiang et al. and future work on energy consumption of the mobile devices used while search will be investigated [27, 28]. *Wireless sensing network:* According to Zhu et al., the current solution is a secure monitoring system that monitors data transmission through controlling effective collection and communication of sensed data which in turn promotes real-time target identification[29]. For real-time object recognition and image detection wireless cameras are used This approach is limited by its inability to handle more complex situations that involve multiple sensing of elements, although it is valuable for encouraging secure monitoring[30].

Meena et al., used an effective algorithm for providing energy-efficient secure transmission of data for secure monitoring [16]. They offered security in data transmission with less consumption of energy using an FLSO protocol. However, there is no indication of the cost of this solution which needs to be established through real-time implementation. The research of Wang et al is valuable for providing energy-efficient secure transmission of data [31]. For future work, more expansion and its implementation in the healthcare sector are performed. *Password-based techniques:* Wu has utilized smart card verification techniques and password-based authentication approaches to ensure the authentication of wireless sensor networks [32]. Password-based techniques and smart-card-based verification are used, which provide secure and authenticated access to the patient's data. However, this research has not addressed usability aspects that are essential for the satisfaction of patients. Nevertheless, this work has contributed to the secure authentication and adequate verification of cryptographic protocols[33]. *Genetic grey wolf optimizer (GGWO) algorithm :* Sujatha et al.developed a reliable fusion scheme through effectively grouping decision-related visual data for all optimized features [34]. Their aim was multi-focus image fusion with the use of a noisy feature removal scheme. However, Poisson noise was not effectively eradicated from the images and, thus, the scheme is suitable only for the creation of an image with brief visual data. *Energy-efficient Fuzzy Vault-based-Security (EFVSM) :* Pirbhudal et al. developed an energy-efficient fuzzy vault-based biometric security approach with the aim of increasing security and authenticity [17]. The biometric-based security mechanism enhances security and the EFVSM reduces energy usage through time-domain analysis using EI. This solution has not secured communication necessary for data consistency and secure transmission [35]. *Wireless smart camera network:* Abas et al.focused on energy- and cost-efficiency and real-time security of the SlugCam, a wireless camera network that facilitates efficient monitoring of videos and does not require regular maintenance [36]. There are issues with data management and storage which can be improved through cloud technology. The plan is to use the SlugCam within the cloud environment for efficient data storage and management [37]. *A heuristic approach to network lifetime maximization:* Singh et al.extended the lifetime of the wireless camera sensor networks through manipulation of sensing angle, viewing angle, and an increase in the number of sensors. The network lifetime in WCSNs is maximized. A limitation is a shortage of energy and storage for sensor data management [38]. Adame et al.used CUIDATS hybrid monitoring system for efficient tracking, allowing real-time health monitoring [39].

## The State of The Art

The biometric technique was utilized in this research which analyses the security parameter which identifies the insecurity caused in the healthcare that results improve in authentication[40]. The work is classified based on fingerprint biometric authentication, face-based access control policies and using a digital chip card. These techniques can be identifying the authentication issue in healthcare applications quickly in a precise manner. However, in state-of-the-art technology, there is a limitation which is not authenticating the healthcare application effectively, and also, it cannot be guaranteed security there [14]. There are the various model is available which can provide the analysis research paper which contains the element that solves the limitation found in state of the art [41]. The above two systems which proposes that solve the limitation of state of the art is shown below in Fig. 1[42]

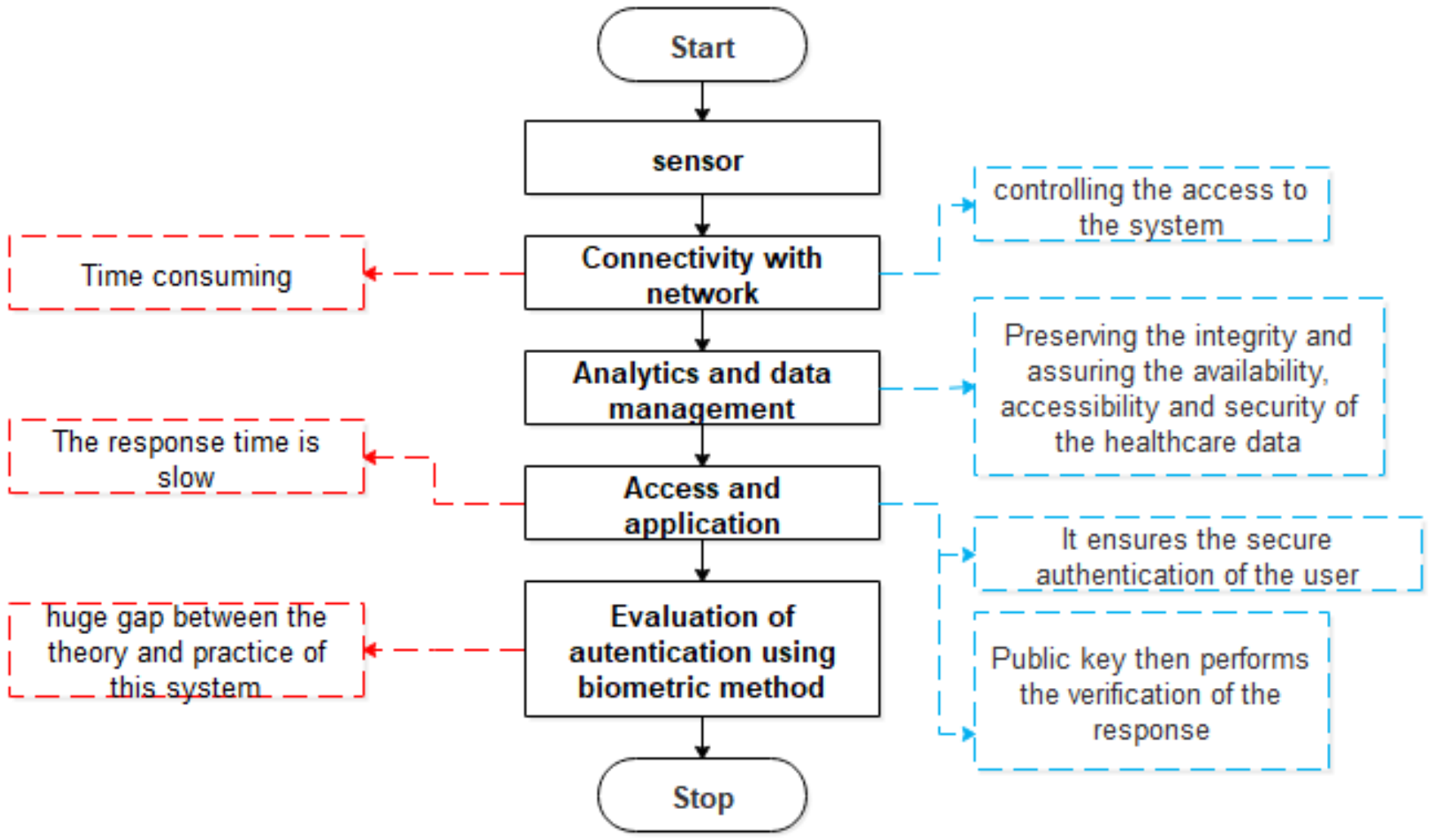

**Fig. 1: State of the art model [42]**

Several methods of biometric authentication-based access control were proposed by this group of researchers, but the work has not resulted in a unified approach required to enhance security, privacy, and control rate[43]. Medical professionals and patients are responsible for maintaining data records and running samples in many of the systems[44]. To overcome these limitations, multiple types of data are processed simultaneously, with details provided in Fig. 2[45].

The Slug cam is used in the healthcare monitoring application for the improvement of authentication using a biometric technique. It provides point-to-point communication in the monitoring application with the camera collecting the healthcare data from the environment and transferring it directly to the requesting medical professional. The Slug cam is a type of wireless sensor camera network that collects data from the healthcare environment data to facilitate monitoring and authentication by improving the communication of data point to point. A smart camera sensor network is used in the data collection process. The smart camera sensors are embedded in every activity of data collection. This is the first step of the proposed system

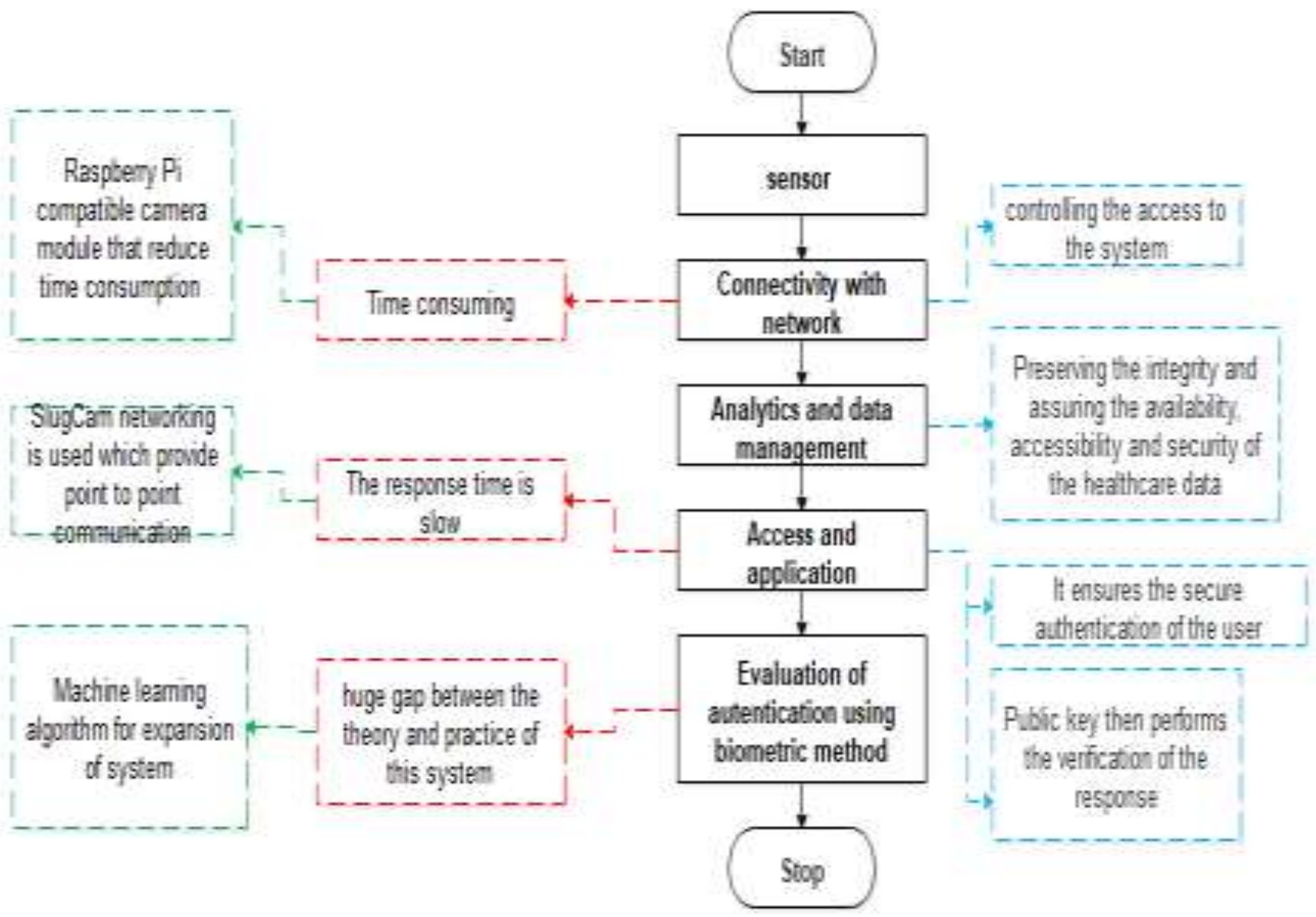

**Fig. 2: Proposed model**

## System components

CNA is based on access control security that facilitates authenticated access to healthcare monitoring applications [5]. With the help of domain specialists in both authenticity and health monitoring, the CNA has been refined to include all factors essential for implementation, identification, and validation[46].

The CAN taxonomy was developed based on a review of current relevant literature. 36 of the initial 208 publications met the following criteria for inclusion.: The research described a complete authentication process for wireless camera sensor networks (WCSNs); used in healthcare monitoring applications even if the concentration was on one of the features only, on authenticity in WCSNs or access control security. Further, the publication described an access control security system that is utilized in the healthcare industry rather than smart homes[47]. 158 results were not included, out of which 23 results were rejected as these publications related to technologies other than those discussed above. several papers only focused on the effectiveness of healthcare monitoring applications without using authentication-based access control security in WCSNs, and the remaining 8 publications were written in foreign languages[48].

On the basis of prior knowledge and information of the healthcare industry, for access control security in WCSNs to guarantee the authenticity of healthcare data, three essentials should be noted: 1) What is the available information that should be collected and identified, 2) how can secure network communication be guaranteed, and 3) how can secured access control and authenticity be guaranteed? A healthcare monitoring system related to authentication-based access control security must be based on 3 essential factors: collection, network communication, and authentication[49].

The first factor used in the CNA taxonomy is a collection that includes the collection of data from WCSNs and then making a comparison with existing needs; the collected data facilitates the identification of intrusion attacks[50]. The properties of its classes involve systems, analysis, processes, etc.[51].

Secondly, the process is network communication, which involves the security of network communication and secure transmission of information utilized for classification. Lastly, the classification is based on authentication, the final product that facilitates managing access control security for health monitoring applications. These three components, their sub-components, and the relationship among them are represented in Table 1[52].

**Table 1: Main attributes and common instances of CNA taxonomy**

| Factor/Class | Main Attributes | General Instances |
|---|---|---|
| **Collection** | Process | Information gathering |
| **Camera** | Device | Raspberry Pi compatible camera module, Sparse camera network-based video surveillance system |
| **Data pre-processing** | Process | Wearable technology, ECG Signals, wireless sensors |
| **Feature extraction** | Acquisition sensors | Personal sensor networks (PSNs), Body sensor networks (BSNs), Multimedia devices (MDs), Template-based approach |
| **Network Communication** | Underlying Process | Secure data transmission |
| **Line configuration** | Services | Point to point, multi-point |
| **Biometric** | Implementation | SlugCam networking, SigQuality software |
| **Connectivity** | Connection Type | 3G/4G, Wi-Fi, ZigBee, Wired |
| **Authentication** | System | The patient health monitoring system |
| **Activity recognition** | Underlying process | Context-aware healthcare services, Visual information analysis, |
| **Access control model** | System | Biometric technology, surveillance algorithm, one/two/three-factor authentication scheme |
| **Healthcare application monitoring** | Analysis | Remote video monitoring, Efficient fuzzy vault-based security method (EFVSM) |

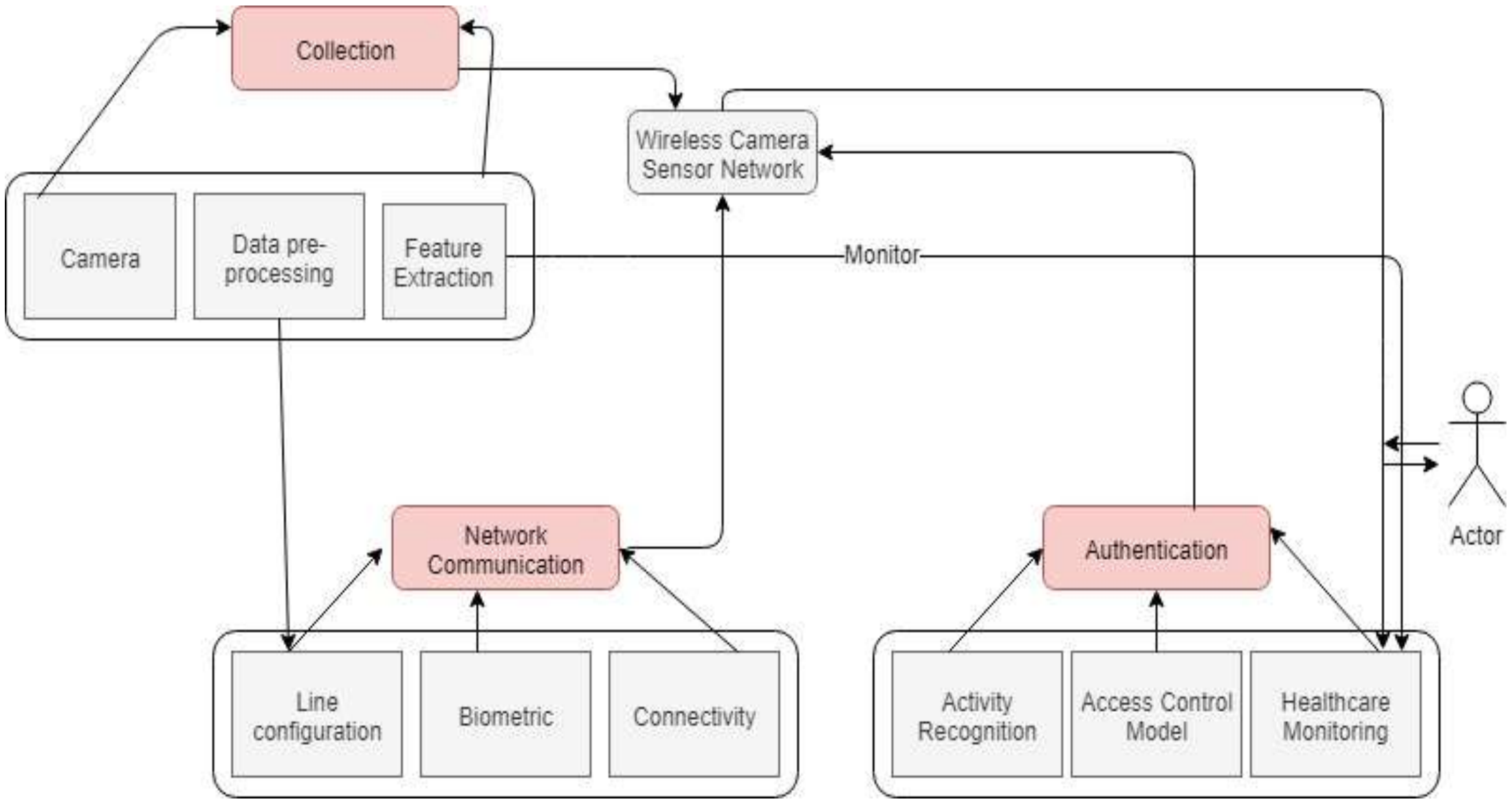

**Fig. 3: Three-factors of the authentication-based access control taxonomy**

The three factors of the authentication-based access control taxonomy for healthcare applications are Collection, Network communication, and Authentication, represented in Fig. 3. Table 1 depicts the components and classes and their connection with one another.

The remaining segment involves the definition of three factors, collection, network communication, and authentication, as well as their subclasses. The justification for the inclusion of these factors is based on their function of *effective data collection, secure network communication and restriction of access control to authorized users* only; therefore, these are utilized for the purpose of classification[53]. Diagrams are provided of the classes and sub-classes which develop each of the factors[54]. The three most significant subclasses of Collection are a camera, data pre-processing, feature extraction [55]. The camera is the device that facilitates monitoring healthcare applications. It is responsible for gathering data with the help of acquisition sensors and processing information - represented in Fig. 4[56]

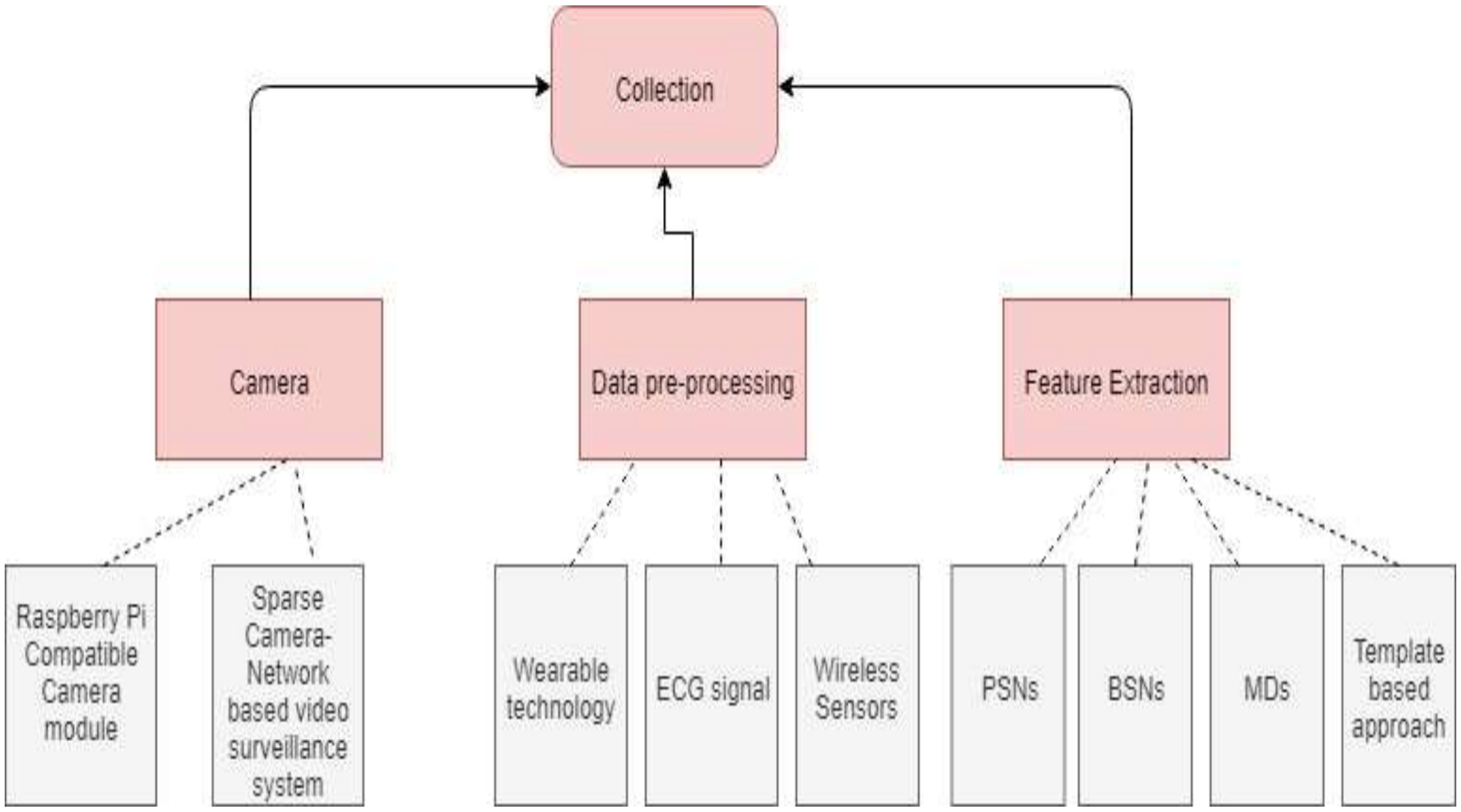

Fig. 4: Collection class and their sub-classes

*Network Communication:* The subclasses of Network communication are line configuration, biometrics and connectivity[57]. They facilitate secure data transmission using different connectivity for transmitting the information to data centers where it can be accessed by authorized users - represented in Fig. 5[58].

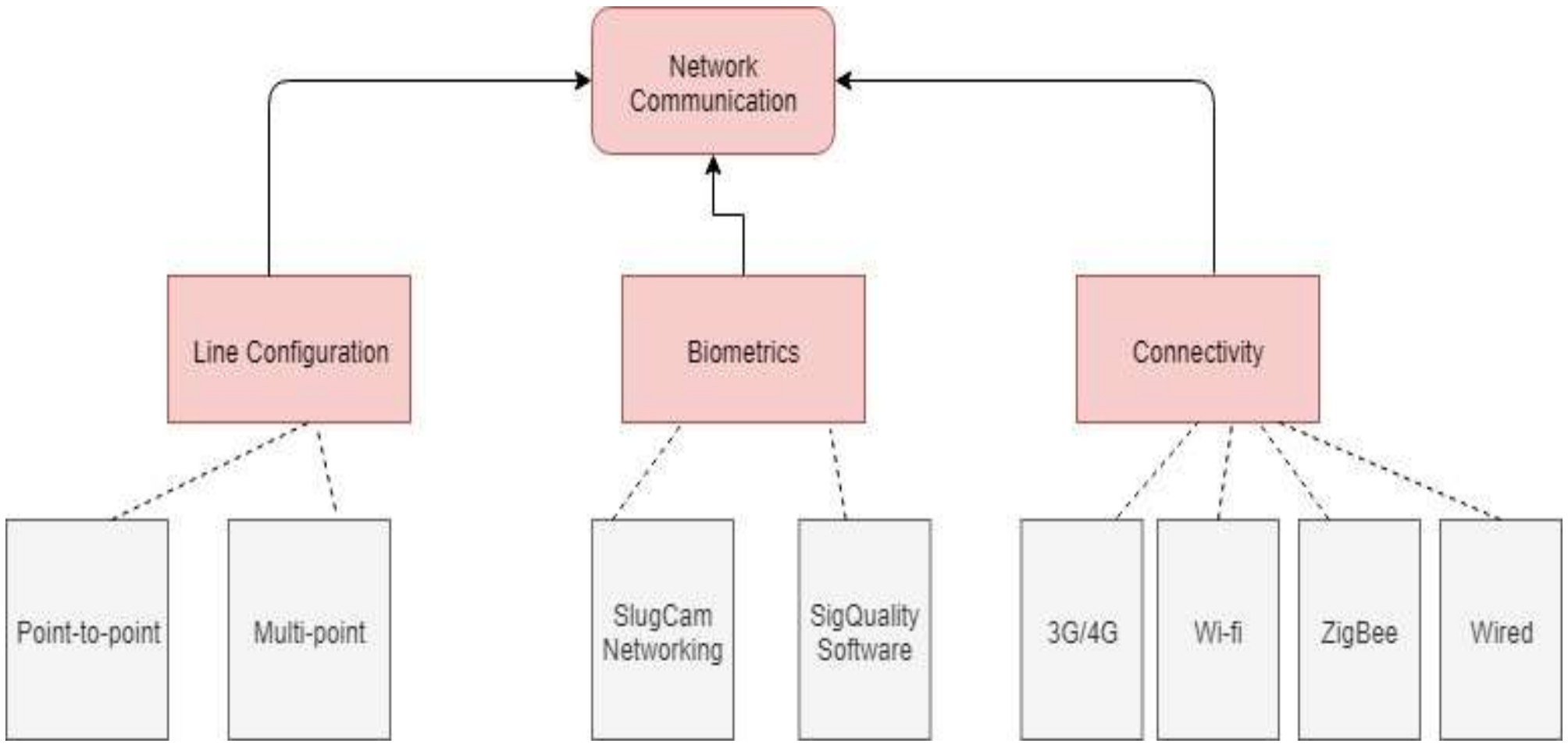

Fig. 5: Network Communication class and their sub-classes

*Authentication:* The subclasses of Authentication are activity recognition, access control and health monitoring[59]. They are used for ensuring the effective detection of intrusions through recognition of activity, access control security, and effective monitoring. They manage access control security of the healthcare application, which is represented in Fig. 6 [11]
.

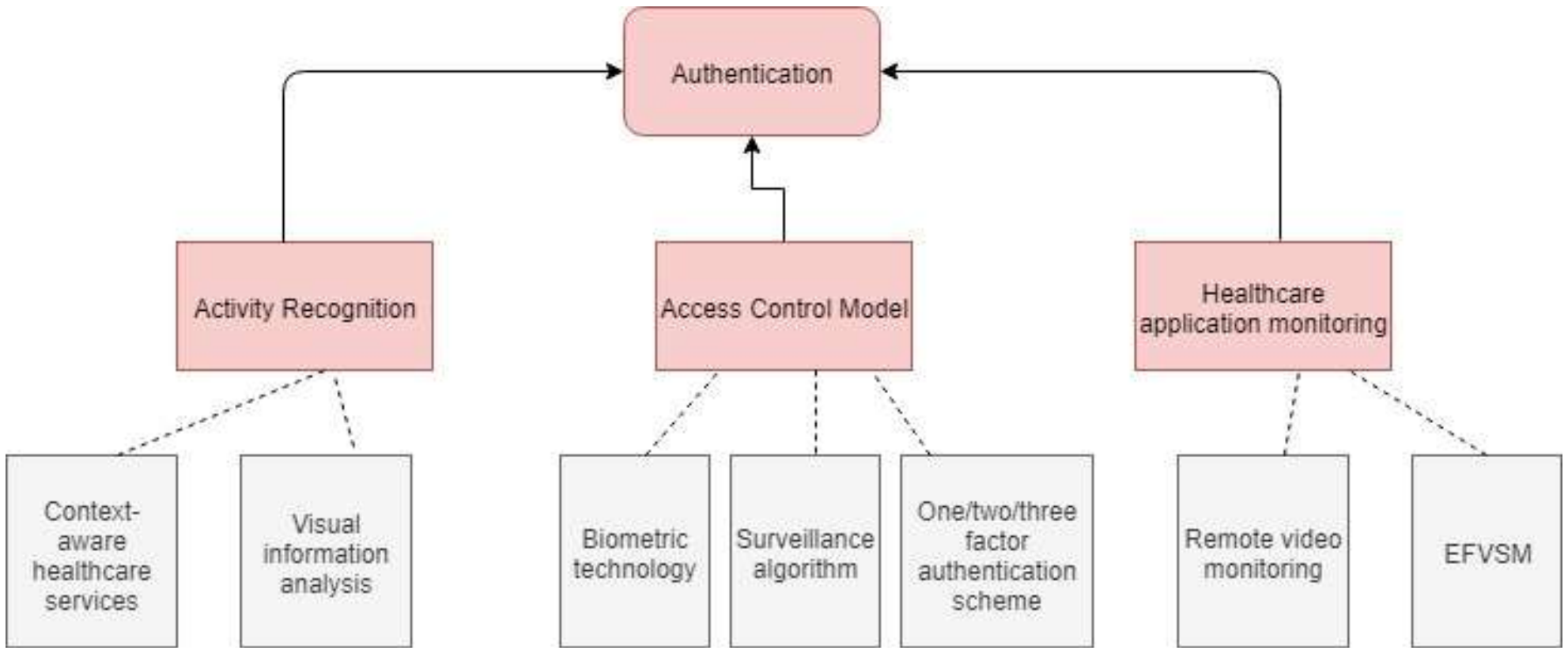

Fig. 6: Authentication class and their sub-classes

Collection of data is a significant part of the classification and is highlighted to stress the significance of the acquisition of data through the integration of camera sensors on the human body. It recognizes and characterizes intrusions that are also communicated to the end-user. Thus, the effective collection of data enhances the detection of intrusion attacks and manages the monitoring of health-related data[60].

The problem is how to safeguard the system from intrusion[61]. Unauthorized access by any individual is not desirable as insecure network communication can lead to the loss of data or reduce packet delivery ratio, which in turn creates difficulties in the of identification of intrusions[62]. Furthermore, the efficient aggregation, classification, and analysis of collected data is required for the effectiveness of the healthcare monitoring applications[63]. Therefore, it is significant to effectively analyze the collected data, secure the transmission of information via a network and emphasize access control security of the health monitoring application to enhance authenticity [64].

## System classification

Existing systems for access control to healthcare data are the basis on which systems classification was carried out. Existing relevant literature was analyzed with the initial search returning 208 potential systems. Only 30 of these met the inclusion criteria [65]. These were currency (2017 onward [6] and publication in quality (Q1 or Q2)[66], and relevance in terms of access control security in the healthcare sector[67]and network communication and authentication in the healthcare sector[68]. Details of the selected systems are shown in Table 2, followed by analysis in Tables 3-5.

Table 2 Classification of the research work on the basis of system primary components

| S.No. | Reference | Collection | | | Network communication | | | Authentication | | |
|---|---|---|---|---|---|---|---|---|---|---|
| | | Camera | Data pre-processing | Feature extraction | Line configuration | Biometric | Connectivity | Activity recognition | Access control model | Healthcare application monitoring |
| 1. | (Abas, et al., 2018) | Raspberry Pi compatible camera module | Wearable technology | Multimedia devices (MDs), | Point to point | SlugCam networking | Not available | Context-aware healthcare services | Biometric technology | Remote video monitoring |
| 2. | (Hamidi, 2019) | Video camera | wireless sensors | Not available | Point to point | Face Recognition | Wired | Context-aware healthcare services | surveillance algorithm | Remote video monitoring |
| 3. | (Mshali, et al., 2018) | Network architecture for cloud-based WBANs | Not available | Haar wavelets. | Point to point | SlugCam networking | Wired | Context-aware healthcare services | one/two/three-factor authentication scheme | Remote video monitoring |
| **4.** | (Chang, et al., 2018) | Video camera | Wearable technology | Template-based approach | Not available | Finger Geometry Recognition | Not available | Context-aware healthcare services | Biometric technology | Remote video monitoring |
| **5.** | (Liu & Chung, 2017) | Network architecture for cloud-based WBANs | Not available | Haar wavelets. | Star topology | SlugCam networking | 3G/4G | Context-aware healthcare services | Biometric technology | Not available |
| **6.** | Shakil, et al., 2017) | Video camera | wireless sensors | Multimedia devices (MDs), | Not available | SlugCam networking | Wired | Context-aware healthcare services | Biometric technology | Remote video monitoring |
| **7.** | (Nandhini & Radha, 2017) | Video camera | Not available | Template-based approach | Point to point | Face Recognition | Not available | Context-aware healthcare services | one/two/three-factor authentication scheme | Remote video monitoring |
| **8.** | (Challa, et al., 2018) | Raspberry Pi compatible camera module | Not available | Personal sensor networks (PSNs), | Not available | SigQuality software | 3G/4G | Context-aware healthcare services | one/two/three-factor authentication scheme | Not available |
| **9.** | (Jiang, 2018) | Video camera | | Template-based approach | Star topology | SlugCam networking | Wired | Context-aware healthcare services | biometric | Remote video monitoring |
| **10.** | (Wang, et al., 2018) | Network architecture for cloud-based WBANs | Wearable technology | Not available | Point to point | Finger Geometry Recognition | 3G/4G | Context-aware healthcare services | one/two/three-factor authentication scheme | Remote video monitoring |
| **11.** | (Wu, et al., 2018) | Video camera | Not available | Color histograms | Point to point | Face Recognition | Wired | Context-aware healthcare services | one/two/three-factor authentication scheme | Remote video monitoring |
| **12.** | (Wu, et al., 2017) | Video camera | Not available | Personal sensor networks (PSNs), | Point to point | Finger Geometry Recognition | Not available | Context-aware healthcare services | biometric | Remote video monitoring |
| **13.** | (Zhu, et al., 2018) | Not available | Not available | Color histograms | multi-point | Face Recognition | Wi-Fi, ZigBee | Context-aware healthcare services | Biometric technology | Remote video monitoring |
| **14.** | (Meena & Sharma, 2018) | Network architecture for cloud-based WBANs | wireless sensors | Haar wavelets. | multi-point | Face Recognition | Wired | Context-aware healthcare services | encryption | Remote video monitoring |
| **15.** | (Amin, et al., 2018) | Video camera | Not available | Template-based approach | multi-point | SigQuality software | Not available | Context-aware healthcare services | key | Remote video monitoring |
| **16.** | (Wazid, et al. 2018) | Not available | Not available | Haar wavelets. | Not available | Face Recognition | Wi-Fi, ZigBee | Visual information analysis | key | Not available |
| **17.** | (Singh, et al. 2017) | Network architecture for cloud-based WBANs | Wearable technology | Body sensor networks (BSNs), | multi-point | Face Recognition | Wi-Fi, ZigBee | Visual information analysis | Biometric technology | Remote video monitoring |
| **18.** | (Sujatha & Punithavathani, 2018) | Not available | ECG Signals | Haar wavelets. | Star topology | Not available | 3G/4G | Not available | surveillance algorithm | Remote video monitoring |
| **19.** | (Wang, et al. 2017) | Video camera | Not available | Haar wavelets. | multi-point | SigQuality | Wi-Fi, ZigBee | Visual informati | Encryption | Efficient fuzzy vault- |
| | | | | | | software | | on analysis | | based security method (EFVSM) |
| **20.** | (Wang, et al., 2018) | Surveillance camera | ECG Signals | Body sensor networks (BSNs), | multi-point | Finger Geometry Recognition | Not available | Visual information analysis | encryption | Not available |
| **21.** | (Pirbhulal, et al., 2018) | Not available | Wearable technology | Haar wavelets. | Not available | Face Recognition | Wi-Fi, ZigBee | Visual information analysis | Biometric technology | Efficient fuzzy vault-based security method (EFVSM) |
| **22.** | (Jiang, et al., 2017) | Network architecture for cloud-based WBANs | Wearable technology | Body sensor networks (BSNs), | multi-point | SigQuality software | Wi-Fi, ZigBee | Visual information analysis | encryption | Remote video monitoring |
| **23.** | (Hossain, et al., 2018) | Surveillance camera | Not available | Haar wavelets. | Star topology | Not available | 3G/4G | Visual information analysis | Key | Efficient fuzzy vault-based security method (EFVSM) |
| **24.** | (Punj & Kumar, 2018) | Network architecture for cloud-based WBANs | ECG Signals | Template-based approach | multi-point | Finger Geometry Recognition | 3G/4G | Visual information analysis | key | Not available |
| **25.** | (Hamidi, 2019) | Raspberry Pi compatible camera module | wireless sensors | Haar wavelets. | Not available | SigQuality software | 3G/4G | Not available | Biometric technology | Efficient fuzzy vault-based security method (EFVSM) |
| **26.** | (Das, et al., 2018) | Not available | Not available | Body sensor networks (BSNs), | multi-point | Finger Geometry Recognition | Wi-Fi, ZigBee | Visual information analysis | surveillance algorithm | Efficient fuzzy vault-based security method (EFVSM) |
| **27.** | (Adame, et al., 2018) | Sparse camera network-based video surveillance system | ECG Signals | Template-based approach | Star topology | Not available | Wi-Fi, ZigBee | Visual information analysis | Biometric technology | Efficient fuzzy vault-based security method (EFVSM) |
| **28.** | (Dodangeh & Jahangir, 2018) | Not available | wireless sensors | Multimedia devices (MDs), | Not available | SigQuality software | Wi-Fi, ZigBee | Visual information analysis | key | Not available |
| **29.** | (Zhou, et al., 2019) | Raspberry Pi compatible camera module | wireless sensors | Template-based approach | Star topology | Finger Geometry Recognition | 3G/4G | Visual information analysis | surveillance algorithm | Efficient fuzzy vault-based security method (EFVSM) |
| **30.** | (Lwamo, et al., 2019) | Sparse camera network-based video surveillance system | ECG Signals | Multimedia devices (MDs), | Star topology | Not available | Wi-Fi, ZigBee | Visual information analysis | Biometric technology | Not available |

**Table 3 Classification of the research work on the basis of collection component attributes**

| S.No. | Reference | Collection | | | | | | | | |
|---|---|---|---|---|---|---|---|---|---|---|
| | | Camera | | Data pre-processing | | | Feature extraction | | | |
| | | Raspberry Pi compatible camera module | Sparse camera network-based video surveillance system | Wearable technology | ECG Signals | wireless sensors | Personal sensor networks (PSNs) | Body sensor networks (BSNs) | Multimedia devices (MDs) | Template based approach |
| **1.** | (Abas, et al., 2018) | ✓ | | | | | | ✓ | | |
| **2.** | (Hamidi, 2019) | ✓ | | | ✓ | ✓ | ✓ | | | |
| **3.** | (Mshali, et al., 2018) | | ✓ | | ✓ | ✓ | | | ✓ | |
| **4.** | (Chang, et al., 2018) | | ✓ | ✓ | | ✓ | | | ✓ | ✓ |
| **5.** | (Liu & Chung, 2017) | | ✓ | ✓ | | | | | ✓ | |
| **6.** | Shakil, et al., 2017) | | ✓ | ✓ | | | | ✓ | | |
| **7.** | (Nandhini & Radha, 2017) | | ✓ | | ✓ | ✓ | ✓ | | | ✓ |
| **8.** | (Challa, et al., 2018) | ✓ | | | | ✓ | | | ✓ | |
| **9.** | (Jiang, 2018) | | ✓ | ✓ | | | | ✓ | | |
| **10.** | (Wang, et al., 2018) | | ✓ | | | ✓ | | ✓ | | ✓ |
| **11.** | (Wu, et al., 2018) | | ✓ | | | | | ✓ | | |
| **12.** | (Wu, et al., 2017) | ✓ | | ✓ | ✓ | | ✓ | | | |
| **13.** | (Zhu, et al., 2018) | ✓ | | ✓ | ✓ | | | | ✓ | |
| **14.** | (Meena & Sharma, 2018) | ✓ | | ✓ | ✓ | | | | | ✓ |
| **15.** | (Amin, et al., 2018) | | | ✓ | ✓ | | | | ✓ | |
| **16.** | (Wazid, et al. 2018) | | ✓ | | | ✓ | | ✓ | | |
| **17.** | (Singh, et al. 2017) | | ✓ | | | ✓ | ✓ | | | |
| **18.** | (Sujatha & Punithavathani, 2018) | ✓ | | | | ✓ | | | | ✓ |
| **19.** | (Wang, et al. 2017) | ✓ | | ✓ | | | | | | ✓ |
| **20.** | (Wang, et al., 2018) | | ✓ | | | ✓ | ✓ | | | ✓ |
| **21.** | (Pirbhulal, et al., 2018) | ✓ | | ✓ | ✓ | | ✓ | | | ✓ |
| **22.** | (Jiang, et al., 2017) | ✓ | | ✓ | ✓ | | ✓ | | | ✓ |
| **23.** | (Hossain, et al., 2018) | ✓ | | | | ✓ | ✓ | | ✓ | |
| **24.** | (Punj & | ✓ | | | | ✓ | ✓ | | ✓ | |
| | Kumar, 2018) | | | | | | | | | |
| **25.** | (Hamidi, 2019) | ✓ | | | ✓ | ✓ | | | ✓ | |
| **26.** | (Das, et al., 2018) | ✓ | | | ✓ | ✓ | | | ✓ | |
| **27.** | (Adame, et al., 2018) | ✓ | | | ✓ | | | ✓ | | |
| **28.** | (Dodangeh & Jahangir, 2018) | | ✓ | ✓ | | ✓ | | | ✓ | |
| **29.** | (Zhou, et al., 2019) | ✓ | | | | | ✓ | | ✓ | |
| **30.** | (Lwamo, et al., 2019) | | ✓ | ✓ | | ✓ | | | | ✓ |

**Table 4 Classification of the research work on the basis of network communication attributes**

| **S.No.** | **Reference** | **Network communication** | | | | | | | | **Implementation step** |
|---|---|---|---|---|---|---|---|---|---|---|
| | | **Line configuration** | | **Biometric** | | **Connectivity** | | | | |
| | | **Point to point** | **Multi point** | **SlugCam networking** | **SigQuality software** | **Wired** | **Wi-Fi** | **Zigbee** | **3G/4G** | Intraoperative |
| **1.** | (Abas, et al., 2018) | ✓ | | ✓ | | ✓ | | ✓ | | Intraoperative |
| **2.** | (Hamidi, 2019) | ✓ | | ✓ | | ✓ | | ✓ | | Intraoperative |
| **3.** | (Mshali, et al., 2018) | ✓ | | ✓ | | ✓ | | ✓ | ✓ | Intraoperative |
| **4.** | (Chang, et al., 2018) | ✓ | | | ✓ | ✓ | | ✓ | ✓ | Intraoperative |
| **5.** | (Liu & Chung, 2017) | ✓ | | | ✓ | | ✓ | | ✓ | Intraoperative |
| **6.** | Shakil, et al., 2017) | ✓ | | | ✓ | | ✓ | | ✓ | Intraoperative |
| **7.** | (Nandhini & Radha, 2017) | ✓ | | | ✓ | | ✓ | | | Intraoperative |
| **8.** | (Challa, et al., 2018) | ✓ | | | ✓ | ✓ | | | | Intraoperative |
| **9.** | (Jiang, 2018) | ✓ | | | ✓ | ✓ | | | | Intraoperative |
| **10.** | (Wang, et al., 2018) | ✓ | | | ✓ | | ✓ | | ✓ | Intraoperative |
| **11.** | (Wu, et al., 2018) | ✓ | | ✓ | | ✓ | | ✓ | ✓ | Intraoperative |
| **12.** | (Wu, et al., 2017) | ✓ | | | ✓ | ✓ | ✓ | ✓ | ✓ | Intraoperative |
| **13.** | (Zhu, et al., 2018) | ✓ | | | ✓ | ✓ | | | ✓ | Intraoperative |
| **14.** | (Meena & Sharma, 2018) | ✓ | | | ✓ | | ✓ | | | Intraoperative |
| **15.** | (Amin, et al., 2018) | | ✓ | | ✓ | | ✓ | | ✓ | Intraoperative |
| **16.** | (Wazid, et al. 2018) | | ✓ | | ✓ | | ✓ | | ✓ | Intraoperative |
| **17.** | (Singh, et al. 2017) | | ✓ | ✓ | | | ✓ | | ✓ | Intraoperative |
| **18.** | (Sujatha & Punithavathani, 2018) | | ✓ | | ✓ | ✓ | | ✓ | ✓ | Intraoperative |
| **19.** | (Wang, et al. 2017) | | ✓ | ✓ | | ✓ | ✓ | | ✓ | Intraoperative |
| **20.** | (Wang, et al., | | ✓ | ✓ | | ✓ | | ✓ | ✓ | Intraoperative |
| | 2018) | | | | | | | | | |
| **21.** | (Pirbhulal, et al., 2018) | | ✓ | ✓ | | ✓ | | | ✓ | Intraoperative |
| **22.** | (Jiang, et al., 2017) | ✓ | | | ✓ | | ✓ | | | Intraoperative |
| **23.** | (Hossain, et al., 2018) | | ✓ | | ✓ | | ✓ | | | Intraoperative |
| **24.** | (Punj & Kumar, 2018) | | ✓ | | ✓ | | ✓ | | | Intraoperative |
| **25.** | (Hamidi, 2019) | ✓ | | ✓ | | | ✓ | | | Intraoperative |
| **26.** | (Das, et al., 2018) | | ✓ | ✓ | | ✓ | | | | Intraoperative |
| **27.** | (Adame, et al., 2018) | | ✓ | ✓ | | ✓ | ✓ | | ✓ | Intraoperative |
| **28.** | (Dodangeh & Jahangir, 2018) | | ✓ | | ✓ | ✓ | | | ✓ | Intraoperative |
| **29.** | (Zhou, et al., 2019) | | ✓ | | ✓ | | ✓ | | ✓ | Intraoperative |
| **30.** | (Lwamo, et al., 2019) | ✓ | | | ✓ | | ✓ | | ✓ | Intraoperative |

**Table 5 Classification of the research work on the basis of authentication component attributes**

| S.No. | Reference | Authentication | | | | | | |
|---|---|---|---|---|---|---|---|---|
| | | Activity recognition | | Access control model | | | Healthcare application monitoring | |
| | | Context-aware healthcare services | Visual information analysis | Biometric technology | One/two/three-factor authentication scheme | Surveillance algorithm | Remote video monitoring | Efficient fuzzy vault-based security method (EFVSM) |
| **1.** | (Hamidi, 2019) | ✓ | | ✓ | ✓ | ✓ | | |
| **2.** | (Das, et al., 2018) | ✓ | | ✓ | ✓ | ✓ | | |
| **3.** | (Adame, et al., 2018) | ✓ | | | | | ✓ | ✓ |
| **4.** | (Dodangeh & Jahangir, 2018) | | ✓ | | | | | |
| **5.** | (Zhou, et al., 2019) | ✓ | | ✓ | | ✓ | | ✓ |
| **6.** | (Lwamo, et al., 2019) | | ✓ | | | ✓ | ✓ | |
| **7.** | (Nandhini & Radha, 2017) | | ✓ | ✓ | ✓ | ✓ | | ✓ |
| **8.** | (Challa, et al., 2018) | ✓ | ✓ | ✓ | ✓ | | ✓ | ✓ |
| **9.** | (Jiang, 2018) | | ✓ | ✓ | ✓ | ✓ | ✓ | |
| **10.** | (Wang, et al., 2015) | | ✓ | ✓ | ✓ | ✓ | | |
| **11.** | (Wu, et al., 2018) | | | ✓ | | | | |
| **12.** | (Wu, et al., 2017) | ✓ | | ✓ | | | ✓ | |
| **13.** | (Zhu, et al., 2018) | ✓ | | ✓ | | | ✓ | |
| **14.** | (Meena & Sharma, 2018) | ✓ | | ✓ | | | ✓ | |
| **15.** | (Amin, et al., 2018) | | | | ✓ | | ✓ | |
| **16.** | (Wazid, et al. 2018) | | ✓ | | ✓ | | ✓ | |
| **17.** | (Abas, et al., 2018) | | ✓ | | ✓ | | | ✓ |
| **18.** | (Hamidi, 2019) | | ✓ | ✓ | | | | ✓ |
| **19.** | (Mshali, et al., 2018) | | ✓ | | | ✓ | | ✓ |
| **20.** | (Chang, et al., 2018) | | ✓ | | | ✓ | ✓ | |
| **21.** | (Liu & Chung, 2017) | ✓ | | | | ✓ | ✓ | |
| **22.** | (Shakil, et al., 2017) | | ✓ | ✓ | | | ✓ | |
| **23.** | (Singh, et al. 2017) | ✓ | | | ✓ | | | ✓ |
| **24.** | (Sujatha, K., & Punithavathani, 2018) | | ✓ | | ✓ | ✓ | | |
| **25.** | (Wang, et al. 2018) | ✓ | | | ✓ | ✓ | | |
| **26.** | (Wang, et al., 2017) | ✓ | | | ✓ | | ✓ | |
| **27.** | (Pirbhulal, et al., 2018) | | ✓ | | | | | |
| **28.** | (Jiang, et al., 2017) | ✓ | | | | | | |
| **29.** | (Hossain, et al., | ✓ | | ✓ | | ✓ | ✓ | |
| | 2018) | | | | | | | |
| **30.** | (Punj & Kumar, 2018) | ✓ | ✓ | ✓ | | | | ✓ |

*Collection:* All 30 systems (see above) use pre-processed data [69]collected through sensors and cameras[46] [48].The camera is the primary tool for the collection of data, generally raspberry pi or sparse camera. In the research of Abas et al. [2], visual is processed using a raspberry pi device with slugcam, while multi-camera networks are used in healthcare for the identification of complex events [2].
*Data pre-processing:* The systems were developed for access control in healthcare through monitoring in which pre-processing plays an important role. In the research of Yao et al[71], a biometric technique is used for security enhancement through which physical characteristics are identified using pre-processing of the image. Zhou et al. [30] tracked real-time objects using a wireless camera for image detection and pre-processing [74].
*Feature extraction:* Feature extraction is another major sub-component. Shakil et [15]extracted features using SigQuality software, which checks signatures, while Sujatha et al. used a noisy feature removal scheme [34].
*Network communication:* Network communication is the second major component of the CNA taxonomy which facilitates the intraoperative environment of the system and all systems rely on a network.
*Line configuration:* This is connecting system components with each other for transmission and receiving purposes. In-line security configuration is the a key attribute of security. In the research of Jiang et al., a point-to-point authentication protocol is used for access authorization that will provide security for health data [76].
*Biometrics:* Biometrics have been a focus for some systems, as for Hamidi et al. [14] and Abas et al. ho used Slugcam networking [36].
*Connectivity:* Connectivity is a significant component of overall systems security, facilitating communication between a data center and an end-user. Zhu et al. [31] used Wi-Fi-based wireless transmission to detect real-time objects[29].
The primary goal of network communication is to enhance the security of a network using biometrics and line configuration. If the connection between users is secure, overall security is enhanced [8].
*Authentication:* Authentication is the third major component of the CCA taxonomy. It provides detection intrusion through the identification of activities and facilitates access control[27].
*Activity recognition:* Activity recognition is the approach in which the user will identify the activities that will occur in healthcare. In the research of the consistency is attained using the energy-efficient fuzzy vault-based security method [17]. A noisy feature removal scheme is also be used in the context-aware healthcare service, which provides the visual information analysis to the data[34].
*Access control:* Access control is commonly used in healthcare to secure sensitive data. It allows access by authorized users only. Surveillance algorithms provide a defensive barrier using k-coverage[35].

## System Components Validation and Evaluation

Validation demonstrates that a system has added value. All 30 publications here reviewed, evaluated, and validated their proposed system to some extent. Thus, most papers focused on either accuracy of security-based authentication techniques using cameras or efficiency (Table 6).

**Table 6- Validation and the evaluation of Healthcare monitoring applications based on WCSNs**

| S.No. | Authors | Area/Domain | Validated or evaluated components | Criteria of study | Evaluation and validation method | Output results |
|---|---|---|---|---|---|---|
| 1. | (Abas, et al., 2018) | Outdoor video monitoring | SlugCam networking | • Slugcam software implementation<br>• SlugCam networking<br>• SlugCam validation | SlugCam | The networking functionality, energy-efficiency, onboard processing capabilities, and open system made this an effective approach to be used in a health monitoring system for monitoring the videos of the patients. |
| 2. | (Hamidi, 2019) | Smart Healthcare using IoT | Network server | • Authentication process<br>• Network server<br>• User verification<br>• Web application | Biometric technique | The appropriate investigation is done in this work and proper analysis of biometric security is provided. |
| 3. | (Mshali, et al., 2018) | Smart homes for health monitoring | Data acquisition through sensing | • Data acquisition through Sensing<br>• Network Communicatio n<br>• Data processing and validation | Multimedia devices (MDs) | The data filtering methods and aggregation approaches could have employed for the security purpose. |
| 4. | (Chang, et al., 2018) | IoT applications | BRA Approach implementation | Problem Finding BRA Approach implementation Validation and Performance | Maximum Disjoint Paths (MDP) | Maximum Disjoint Paths (MDP) mechanism is considered to be a centralized optimal solution and the control packets of BRA is slightly reduced. |
| 5. | (Liu & Chung, 2017) | Healthcare sector | Login phase | • Registration phase<br>• Login phase<br>• Authentication phase. | Data transmission mechanism | The account credentials, passwords are same that depicts the authenticity of the users and then access is granted which is not adequate enough. |
| 6. | (Shakil, et al., 2017) | Biometric authentication of e-healthcare data | Pre-processing of the feature dataset | • Pre-processing of the feature dataset<br>• Feature Extraction<br>• Training and Verification | e-healthcare | This solution facilitates in overcoming the shortcoming related to the forgetting passwords and token theft through enhancing the accuracy rate for secure e-healthcare data access. |
| 7. | (Nandhini & Radha, 2017) | Ensuring video privacy in the healthcare industry | Block division | • Extract frames<br>• Block division<br>• Block selection process<br>• Diagonal sum technique | Background subtraction method, Block selection process | the advancement made for reducing the storage needs by 92% and energy consumption by 53%. |
| 8. | (Challa, et al., 2018) | User authentication in the healthcare sector | User revocation and re-registration phase | • The System set-up phase<br>• Sensor node pre-deployment phase<br>• Registration phase<br>• Login-phase<br>• Authentication and key-agreement phase | Key agreement protocol | This research will enhance the level of security with the ECC-based secure three-way authentication method. |
| **9.** | (Jiang, 2018) | Electric industrial context | Integration of low-level features | • Track correspondence modelling<br>• Sparse camera network-based video surveillance system<br>• Global automatic and comprehensive analysis | Ideal sparsity camera network monitoring system | The solutions are not able to analyze the visual cues in the network. |
| **10.** | (Wang, et al., 2018) | Security through efficient target tracking in the healthcare industry | Reducing the moving distance | • Identification of tracking problem<br>• Reducing the moving distance<br>• Managing the target tracking | Wireless camera sensors | The given solution concentrates on managing the sensing range, security and tracking quality. |
| **11.** | (Wu, et al., 2018) | Authentication in wireless sensor networks | Smart-card-based verification | • Password-based authentication<br>• Smart-card-based verification | Password-based techniques, smart-card-based verification | • Effective authentication and verification |
| **12.** | (Wu, et al., 2017) | properties analysis of healthcare applications | Ordering the events | • Verification of the cryptographic protocols<br>• Security testing<br>Ordering the events | Cryptographic protocols | This work performs analysis and testing for the attaining of goal and ensures the verification of the security properties of the cryptographic protocols. |
| **13.** | (Zhu, et al., 2018) | Secure monitoring in the healthcare sector | Target identification and real-time object recognition | • Collection<br>• Communication<br>• Target identification and real-time object recognition | Wireless sensing network | The system is efficient as it ensures accurate detection of images and identification of real-time objects which are used for the secure monitoring of the patients in the field of medical industry. |
| **14.** | (Meena & Sharma, 2018) | Secure data transmission in wireless sensor networks | Clustering process | • Clustering process<br>• Enabling end-to-end encryption<br>Creation of routing table | Clustering process | Security features enhanced. |
| **15.** | (Amin, et al., 2018) | Authentication in wireless medical sensor networks | Medical professional and patient registration | • Setup<br>• Medical professional and patient registration<br>• Login and authentication<br>Password change | AVISPA Tool, BAN logic | The study used the cryptographic hash function for the security purpose that takes 0.004ms which is quite less than the 0.1303ms which are consumed by the AES (Asymmetric encryption algorithm). |
| **16.** | (Wazid, et al. 2018) | Authentication in WBANs in the field of health monitoring | User-registration | • user registration<br>• login<br>• authentication & key agreement<br>• password & biometric update<br>• dynamic body sensor addition<br>mobile device revocation. | 3-factor authentication | 3-factor authentication, secure communication, semantic security, and enhanced network performance. |
| **17.** | (Singh, et al. 2017) | Maximizing network lifetime in Wireless camera sensor network | Random deployment of sensors | • Random deployment of sensors<br>• Classification of | Wireless camera sensor networks (WCNNs) | The random sensor deployment that utilizes low energy in the inactive state and its battery lasts longer contributes as essential factors for lifetime |
| | | (WCSNs) | | sensors into subsets<br>Analysis of network lifetime | | maximization of the network. |
| **18.** | (Sujatha, K., & Punithavathani, 2018) | health monitoring | Feature extraction | • Feature extraction<br>• Removal of noisy features<br>Evaluating performance | Genetic grey wolf optimizer (GGWO) algorithm | visual data with high sharp areas and then combines the data for producing an image with improving quality and high-level definition. |
| **19.** | (Wang, et al. 2017) | Effective surveillance using WSNs | Analysis of barrier coverage formation problem | • Analysis of barrier coverage formation problem<br>• Assign mobile sensors<br>• Performance evaluation | Cluster-based directional barrier graph model | The use of greedy movement algorithm helps in dealing with the gap and optimal approach for deployment is selected. |
| **20.** | (Wang, et al., 2017) | WSNs for coverage problem | Model Construction | • Model construction<br>• Objective formalization<br>• Solution design<br>• Result verification | Probabilistic and directional models | It establishes an idea of non-deterministic problems related to coverage with uncertain features. |
| **21.** | (Pirbhulal, et al., 2018) | Data consistency in telehealth monitoring | Gathering of health information | • Gathering of health information<br>• Feature extraction<br>Analysis of performance | Fuzzy vault based biometric approach | the consumption of energy, it is by 1.423J and processing time by 0.168ms with an efficient detection rate of 91.3% |
| **22.** | (Jiang, et al., 2017) | Wearable health monitoring systems (WHMSs) | Login and authentication phase | • Set-up phase<br>• Medical professional and patient registration phase<br>• Login and authentication phase<br>Password update phase | Wearable sensors, key agreement protocol | The solution has prevented security vulnerabilities and enhances security and efficient authenticity. |
| **23.** | (Hossain, et al., 2018) | Medical health records protection | Adoption of Attributed-based Access Control model | • Integration of OpenID standard<br>• Adoption of Attributed-based Access Control model<br>• Design Delegated Context-aware Capability-based Access Control (DCCapBAC) scheme | OpenID standard | The processing of data is improved and the medical sensors are delegated and the delegation approach facilitates in reducing the consumption of energy and provides more security. |
| **24.** | (Punj & Kumar, 2018) | WBANs in health monitoring | Data analysis | • Data collection<br>• Data aggregation<br>Data analysis | Wireless body area networks (WBANs) | The solution is reliable and it provides the secure transmission of data and cluster is formed that Facilitates in local processing and communicate using wireless connections. |
| **25.** | (Hamidi, 2019) | Smart Healthcare using IoT | Web application | • Authentication process<br>• Network server<br>• User verification | Biometric registration | The appropriate investigation is done in this work and proper analysis of biometric security is provided. |
| | | | | • Web application | | |
| **26.** | (Das, et al., 2018) | Authentication in healthcare applications | User Authentication | • User authentication<br>• Device authentication | single-factor authentication scheme | the healthcare data and facilitate in managing secure access through the process of user authentication and device authentication. |
| **27.** | (Adame, et al., 2018) | Monitoring in the healthcare environment | WSN | • CUIDATS Wristband<br>• RFID (Radio-frequency identification)<br>• WSN (Wireless server network)<br>• RFID-WSN integration | Location sensor | The solution provides secure communication and transmission of data. |
| **28.** | (Dodangeh & Jahangir, 2018) | Access control security or authenticity in WBAN | Intra-WBAN Transmissions | Intra-WBAN Transmissions<br>Inter-WBAN modules | biometric security scheme | It is an effective solution as it uses secure key exchange protocol and the access control security in WBAN is improved. |
| **29.** | (Zhou, et al., 2019) | Authentication in wireless sensor networks | Password-based authentication | • Password-based authentication<br>• Smart-card-based verification | Password-based techniques | This work has resulted in reducing the computation time, monitoring time and communication costs that in total contributed to the effectiveness in continuous health monitoring of the patients. |
| **30.** | (Lwamo, et al., 2019) | Healthcare sector | Login phase | • Registration phase<br>• Login phase<br>• Authentication phase. | Wireless Healthcare sensor network (WHSN) | The phases of this work, registration, login and authentication phase ensure that the account credentials, passwords are same that depicts the authenticity of the users and then access is granted which is not adequate enough. |

# Verification of the proposed system

Qualitative and quantitative methods have been used for the evaluation of the proposed model. The validity of authentication has been achieved through several security techniques. System components were compared with existing taxonomies.

*System acceptance:* An overlap test was conducted for CNA systems components to ensure terms and instances also occur in the literature that was analyzed (Table 2). However, the test is to some extent, qualitative since linguistic relationships are primarily dependent on context. The terms which overlap between the CNA and the provided corpus are shown in Table 6.

*Completeness:* To ensure system completeness, essential components and the subcomponents of the state-of-the-art papers were analyzed. Twenty-one papers between 2017 and 2018 developed a system. However, they generally did not consider security. Fig. 7 analyses the use of components and the subcomponents.

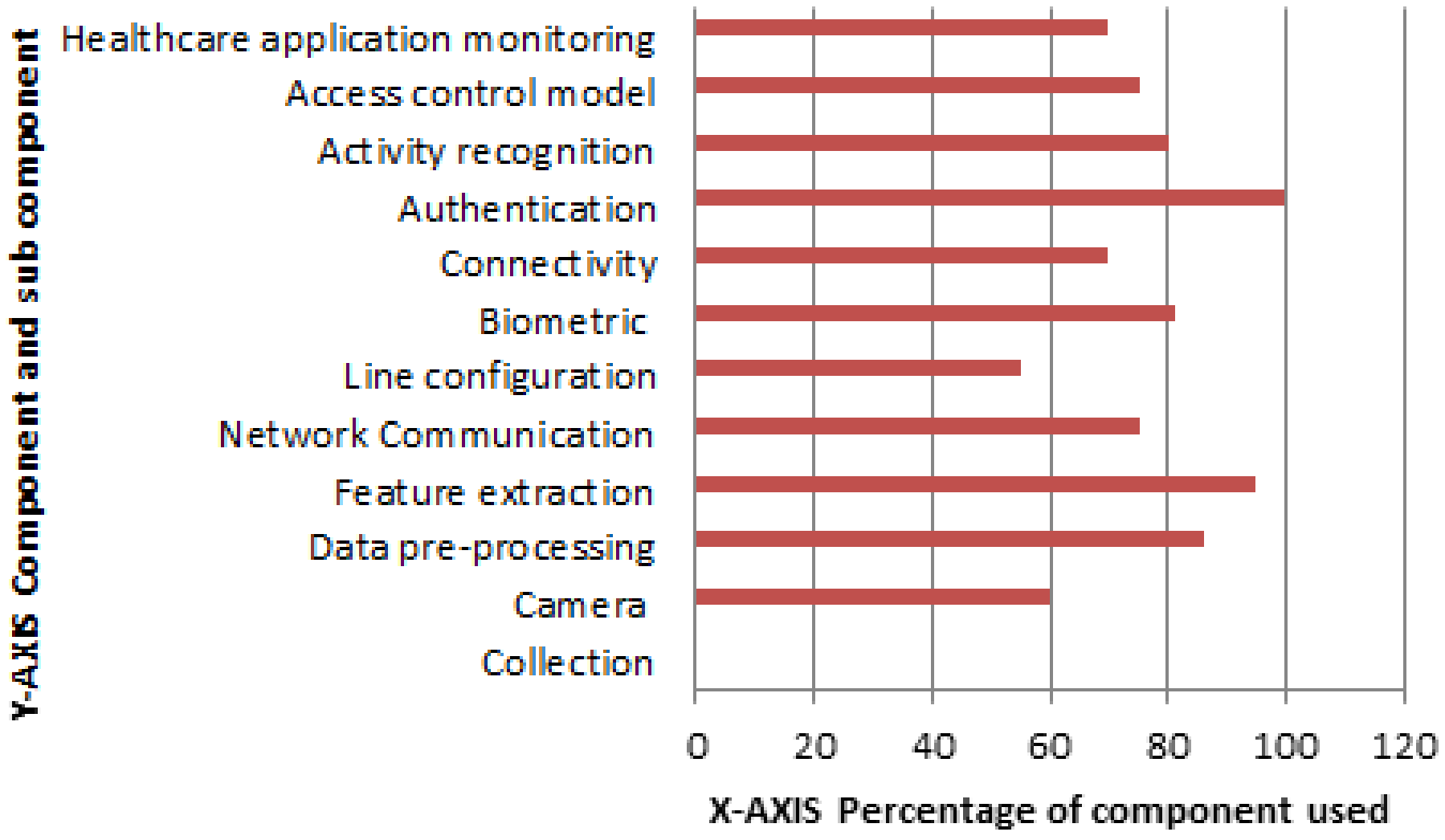

Fig. 7: Existing literature Components and classes

# Discussion

The frequency with which CNA components occur in the literature is analyzed in Table 7 below.

**Table 7: Term frequency table**

| Term | Frequency | Term | Frequency | Term | Frequency |
|---|---|---|---|---|---|
| **Data integration** | 56 | **Camera** | 65 | **Authentication** | 72 |
| **Data mining** | 58 | **Biometric** | 75 | **Connectivity** | 84 |
| **Feature extraction** | 25 | **Line configuration** | 25 | **Access control** | 37 |
| **Activity recognition** | 36 | **Data Collection** | 68 | **Monitoring** | 57 |

In above table 7, the 24 most re-occurring terms are listed with the frequency in the 30 publications used in the research work.

*Collection (Data pre-processing):* Existing systems primarily used cameras and feature extraction in their work. Cameras were used to capture data, and feature extraction was carried out to provide authentication of medical data. Data pre-processing was not included by many researchers, with few of these using the technique for authentication.

*Network communication (Connectivity):* Connectivity was generally not considered, neither for authentication nor privacy of patient data.

*Authentication (Activity recognition):* Few existing systems referred to activity recognition.

*Collection (Wearable technology):* Wearable technology was generally not considered in existing systems. Wireless sensors and ECG signals may have been considered to be more critical.

*Network communication (Wired connectivity):* Wired connectivity is one of the oldest and most reliable technologies for connecting devices and systems, although it may cause a delay in communication and a slow rate of data transmission. Wired connectivity was not found in the existing works, presumably for that reason.

*Authentication (Visual information analysis ):* Visual information analysis is not described by the authors in the research work.

*Collection (Wearable technology):* None of the existing research mentioned wearable technology however, this research suggests that only with the addition of wearable technology can complete data sensing coverage be achieved.

*Network communication (Wired connectivity ):* The discussion section in the research report has proven that the component of the wired connectivity in the transmission and the storage of the collected data is not considered. However, it is known that the inclusion of the wired connectivity into the network will increase the efficiency of the model for transmission of the information.

*Authentication (visual information analysis):* Visual data collected from patients were not considered in the research models.

## Conclusion

This study aimed to provide monitoring and security and privacy of medical data, whereby a server stores the data and access authority lies with medical personnel or patients. Data consists of input from wireless sensors and wearable devices which are processed on the serverA medical professional analysis the data, and remote healthcare monitoring can be provided.

The study provided a comprehensive systematic review of the current approaches and methods used for healthcare monitoring through wireless camera. Additionally, the study identified the limitations of the existing models arise from slow response time due to wired connectivity and the fact that theoretical and experimental results were not consistent. Future work is needed to decrease transmission time, as well as improve data collection methods authentication processes.